\documentclass[prb,aps,amsmath,amssymb,floatfix,prb,twocolumn,showpacs]{revtex4}

\usepackage{epsfig}
\begin{document}

\title{Nonlinear ac conductivity of interacting 1d
electron systems}

\author{Bernd Rosenow and Thomas Nattermann}

\affiliation{Institut f\"ur Theoretische Physik, Universit\"at zu
K\"oln, D-50923 Germany}

\date{September 27, 2005}

\begin{abstract}

We consider low energy charge transport
in one-dimensional (1d)
electron systems with short range interactions under the influence of a 
 random potential.
Combining RG and instanton methods, we calculate the nonlinear ac
conductivity and discuss the crossover between the nonanalytic field
dependence of the electric current at zero frequency and the linear ac
conductivity at small electric fields and finite frequency.

\end{abstract}

\pacs{71.10.Pm, 72.15.Rn, 72.15.Nj}

\maketitle

\section{Introduction}

In 1d electron systems, the effect of both interactions and random
potentials  is very pronounced, and a variety of unusual phenomena
can be observed \cite{Giamarchi,Gruner}. 
The linear \emph{dc} conductivity shows a power law dependence on
temperature $T$ at higher temperatures
\cite{Giamarchi+Schulz,KaneFisher} , but is exponentially small at
low temperatures and vanishes at zero temperature
\cite{Mott,Shklovskii-Efros}.
The \emph{ac} conductivity vanishes like \cite{MoHa68} $\sim
\omega^2 (\ln (1/\omega))^2$ and shows several cross-overs to other
power laws at higher frequencies \cite{Fogler02}.

Much less is known about the \emph{non-linear conductivity}. At zero
temperature and frequency charge transport is only possible by
tunneling of charge carriers, which can be described by instanton
formation. The nonlinear dc-conductivity is characterized by \cite{S73,NaGiDo03,MaNaRo04,FoKe05}
$I \sim \exp( -
\sqrt{E_0 / E}) $  
  provided the system is coupled to a
dissipative bath.  Without such a coupling, the current was recently
suggested to vanish below a critical temperature \cite{GoMiPo05,BaAlAl05}.

In this work, we calculate the low energy {\it non- linear ac
  conductivity} for  systems with random pinning potentials
and discuss the crossover between linear ac response at small fields
and nonlinear dc response at large fields.  To be specific, we
consider a charge density wave (CDW) or spinless Luttinger liquid
(LL) pinned by a random lattice potential which can be described by
the quantum sine-Gordon model with random phases.  We first scale
the system to its correlation length, where the influence of the
potential is strong and a semiclassical instanton calculation
becomes possible. 
 The response is dominated by
energetically low lying two level systems (TLS), whose dynamics is
described by a Bloch equation.

Microfabrication of quantum wires or 1d CDW systems \cite{slot+04}
should allow to test our predictions experimentally. Indeed there is
a number of recent experiments on carbon nanotubes
\cite{TaSuWa00,CuZe04,TzChYi04} and polydiacetylen \cite{AlLeCh04}
which seem to confirm the variable range hopping prediction for the
dc-conductivity made in \cite{NaGiDo03,MaNaRo04}.

\setcounter{equation}{0}

\section{AC conductivity of 1d disordered systems}

In the following, we present a heuristic derivation \cite{Mott}  of the
Mott--Halperin result \cite{MoHa68} for the \emph{ac} conductivity
of a one-dimensional  disordered electron system without
interactions.  In the end of the section, we indicate how this
result can be generalized to interacting electrons.

In one spatial dimension, all electron states are localized and
wave function envelopes decay on the  scale  of the localization
length $\xi_{\rm loc}$.
We divide the system into segments with size $\xi_{loc}$.
The typical energy separation of
 states within one segment is the mean level spacing
$\Delta = 1/(\rho_F \xi_{\rm loc})$,
where $\rho_F$ is the density of states at the Fermi level per unit length.
Levels in neighboring segments are coupled by the Thouless energy
   $t(\xi_{loc}) = \Delta$.  When we consider the
coupling between more distant segments of separation $L$ the
coupling is reduced to $t(L) = \Delta \exp(- L/\xi_{loc})$. The
coupling splits (almost) degenerate energy levels in different
segments by an amount $\Delta E = 2 t(L)$. 
Due to the coupling, the eigenstates of the Hamiltonian are even and odd linear 
combinations of states localized in segments a distance $L$ apart.
A spatially constant ac  electric field $E(t) = E_0 \cos \omega t$
causes transitions between 
levels with a
separation $\Delta E = \hbar \omega$, hence we demand $2 t(L)=\hbar
\omega$ and therefore
%
\begin{equation}
 L_x(\omega) = \xi_{loc}
\ln(2 \Delta / \hbar \; \omega) \ .
\label{tunnellength.eq}
\end{equation}
%
According to Fermi's golden rule, the transition rate for exciting the
even linear combination to the odd one is given by $1/\tau = {2 \pi
\over \hbar}\ {1 \over 4} e_0^2 E_0^2 |\hat{x}_{\epsilon,\epsilon+
\hbar \omega}|^2 \ \rho_F \ \xi_{loc}$. Here, the matrix element
$\hat{x}_{\epsilon,\epsilon+ \hbar \omega}$ of the position operator
has to be calculated between the ground state with energy $\epsilon$
and the excited state with energy $\epsilon + \hbar \omega$, it is
found to be equal to the spatial separation $L_x(\omega)$ of the two
localized states. When calculating the rate of energy absorption ${1
\over 2} \sigma_{\rm ac}(\omega) E_0^2$ due to the excitations of such
two level systems, one has to take into account that each photon
carries the energy $\hbar \omega$, that only transitions from
unoccupied states to occupied states are possible, and that the
occupation probability for a state with energy $\epsilon$ is
determined by the Fermi function $f(\epsilon)$. In this way, one finds
%
\begin{eqnarray}
\sigma_{\rm ac}(\omega) & = & \int d \epsilon \rho_F \hbar \omega f(\epsilon) 
[1 - f(\epsilon)] { 2 \pi \over \hbar}  {e_0^2 \over 2}  L_x(\omega)^2 
\xi_{loc} \rho_F \nonumber \\
&  \approx &  \sigma_0 \  L_x^2(\omega)\ ( \hbar \omega
\rho_F)^2,\,\,\,\  \sigma_0=\frac{e^2_0}{\hbar}\xi_{loc} .
\label{motthalperin.eq}
\end{eqnarray}

 This result can be generalized to an interacting electron system
\cite{FeVi81} by remembering the basic idea of bosonization: the
charge density is defined as the derivative of a displacement field,
and a localized electronic state corresponds to a localized kink in
the displacement field. In addition, the density of states $\rho_F$ at
the Fermi level has to be replaced by the compressibility $\kappa =
{\partial \rho \over \partial \mu}$. With these modifications, the
above derivation can be repeated and one obtains an result analogous
to Eq.~(\ref{motthalperin.eq}).

It is worthwhile to remark that the ac-conductivity
(\ref{motthalperin.eq}) can be rewritten as
%
\begin{equation}
\sigma_{\rm ac} (\omega) \sim \sigma_0 \ e^{-2L_x(\omega)/\xi_{loc}}
\ \left(\frac{L_x(\omega)}{\xi}\right)^2,\,\,\,\ \label{motthalperin.eq2}
\end{equation}
%
where $\xi$ is the correlation length of the system. This result
 resembles the form of the result for the \emph{non-linear
dc}-conductivity
%
\begin{equation}
\sigma_{\rm dc} (E) \sim \sigma_0 \ \frac{\hbar \omega_{\rm
e-ph}}{\Delta} \ e^{-2L_x(E)/\xi_{\rm loc}} \ \left(\frac{L_x(
E)}{\xi}\right)^2 \,\,\,\ \ . \label{quantum_creep.eq}
\end{equation}

Here $L_x(E)=\xi_{\rm loc}\sqrt{\frac{\Delta}{e_0E_0\xi_{\rm loc}}}$
denotes the spatial distance of the energy levels between which the
tunneling events take place, it follows from a variational treatment
\cite{S73,NaGiDo03}. The prefactor $\hbar \omega_{\rm
e-ph}/e_0E_0\xi_{\rm loc}$ takes into account that the dissipation
rate is controlled by the typical frequency for electron phonon
coupling $\omega_{\rm e-ph}$ and that the current is hence of the
order \cite{FoKe05} $\sim e_0\omega_{\rm
e-ph}e^{-2L_x(E)/\xi_{\rm loc}}\,\,\,\,\,$. We note that depending on
the details of the electron phonon coupling, $\omega_{\rm
e-ph}$ may depend on the external electric field. 

Finally, at finite temperatures, Mott variable range hopping gives a
linear \emph{dc}-conductivity which follows from
(\ref{quantum_creep.eq}) by replacing $L_x(E)$ by $L_x
(T)=\sqrt{\Delta/k_BT}$, i.e.\cite{Mott,Teber}
%
\begin{equation}
\sigma_{\rm dc} (T) \sim \sigma_0 \ \frac{\hbar \omega_{\rm
ph}}{\Delta} \ e^{-2L_x(T)/\xi_{\rm loc}} \ \left(\frac{L_x(
T)}{\xi}\right)^2\,\,\,\ \ . \label{vrh.eq}
\end{equation}
%
The cross-over between the three expressions
(\ref{motthalperin.eq2})-(\ref{vrh.eq}) can most easily be
understood by the dominance of a shortest tunneling distance, as
will be discussed further below.
\setcounter{equation}{0}

\section{The Model}


The models we analyze are defined by the euclidean action
%
\begin{eqnarray}
{S\over \hbar} & = & { 1 \over 2 \pi K} \int dx \int_0^{v  \hbar \beta} d y
\left[ \left({\partial \varphi \over \partial x} \right)^2 +  \left({\partial \varphi
\over \partial y}\right)^2
\right.
\label{action.eq}\\
& & \left. - 2 u \cos( p \varphi  + 2 \pi \zeta(x)) + {2 K e_0 \over \pi  v} \varphi  E(y)
\right] \ + {S_{\rm diss} \over \hbar} ,
\nonumber
\end{eqnarray}
%
where we have rescaled time according to $v \tau \to y$, and $\beta =
1/k_B T$.  The dissipative part of the action describes a weak
coupling of the electron system to a dissipative bath, for example
phonons. It is needed for energy relaxation in variable range hopping
processes \cite{MaNaRo04} and for equilibration in the presence of a
strong ac field.  We assume it to be so small that it does not
influence the RG equations for the other model parameters
significantly.  The smooth part of the density is given by ${1 \over
\pi} \partial_x \varphi$, and $p=1,2$ for CDWs and LLs, respectively.
We consider a CDW or LL with Gaussian disorder, which is  described by
 $\zeta(x)$  equally distributed in the
interval $[0, 1]$ with correlation length equal to the lattice spacing
$a$.

For $K > K_c(u)$ the potential is RG irrelevant and decays under the
RG flow, while for $K < K_c(u)$ the potential is relevant and grows,
here \cite{Giamarchi+Schulz} $K_c(0)= 6/p^2$ .  We assume $K <
K_c(u)$ and scale the system to a length $\xi= a e^{l^\ast}$, on
which the potential is strong. After the scaling process, the
parameters $K$, $v$, and $u$ in Eq.~(\ref{action.eq}) are replaced
by the effective, i.e. renormalized but not rescaled, parameters
$K_{\rm eff}$, $v_{\rm
  eff}$, and $u_{\rm eff}$.

 We note that 
 the ratio $K/v $ and hence the
 compressibility $\kappa = {K \over v \hbar \pi} $ is not renormalized
 due to a statistical tilt symmetry \cite{Schultz+88}.  The
 compressibility $\kappa = {\partial \rho \over \partial \mu}$ is used
 as a generalized density of states for interacting systems.  Our
 calculations are valid for energies below
 the generalized mean level spacing $\Delta_{0}
 = {1 \over \kappa \xi}$.

 In this RG calculation, we do not attempt to treat a possible nonlinear
 dependence of coupling parameters on the external electric field.
 The full inclusion of the external field in an equilibrium theory is not possible
 as it renders the ground state of the system  unstable.
 The quantum sine-Gordon model has an infinite number of
 ground states connected by a shift of the phase field by $2 \pi /p$.
 Here, we concentrate on renormalizing each of these
  ground states
 separately and  take into account the coupling between different ground
 states due to the external electric field in the framework of an instanton
 approach.

\setcounter{equation}{0}

\setcounter{equation}{0}

\section{Disordered LL or CDW}


The wall width $1/\sqrt{p^2 u_{\rm eff}} \approx \xi$ of an instanton
solution to the action Eq.~(\ref{action.eq}) is for weak external fields
much smaller than the extension 
of the instanton.
Hence, the instanton action can be expressed in terms of the domain
wall position $X(y)$. 
The discussion of instantons in the case of  random pinning  is
more involved
than e.g. for periodic pinning \cite{Maki77} and the calculation of 
closed form instanton solutions is not possible.
For this reason, 
we look for approximate instanton solutions with a rectangular shape
and extensions $L_x$, $L_y$ in $x$-- and $y$--direction,
respectively.
As the disorder is correlated in time but not in
space, instanton walls in $x$-- and $y$--direction contribute  $L_x
s_y$ and $L_y s_x(x)$ to the action, respectively.  While the surface tension $s_y
= {2 \pi \over p^2 \xi K_{\rm eff}}$ is essentially constant,
the surface tension $s_x$ has a strong and random position
dependence. To calculate the statistical properties of $s_x$, we
make use of the exact solution \cite{Glatz01} of the classical
ground state  of a LL or CDW with random pinning in the following.

In the limit $K_{\rm eff} \ll1$, quantum fluctuations are strongly
suppressed and the (classical) ground state of the model
Eq.~(\ref{action.eq}) can be determined exactly \cite{NaGiDo03}.
After renormalization to the scale $\xi$, the effective action can be
rewritten as a discrete model on a lattice with grid size $\xi$, and
the integration over $x$ can be replaced by a summation over discrete
lattice sites $i = x/\xi$.  In the classical ground state, the
solution $\varphi(x,y)$ does not depend on $y$ any more and the
$y$--integral in Eq.~(\ref{action.eq}) simply yields an overall factor
$v_{\rm eff} \hbar \beta$. Dividing the action by $\hbar \beta$, one
obtains the classical Hamiltonian \cite{NaGiDo03}
%
\begin{equation}
H_{\rm class} = { \Delta_0 \over 2 \pi^2 } \sum_{i=1}^{L_0/\xi}
[(\phi_{i+1}\!  -\!  \phi_i)^2 - 2  \xi^2  u_{\rm eff}
  \cos(p \phi_i\! -
 \! 2 \pi  \zeta_i)] \; .
\label{eq.hamiltonian}
\end{equation}
%
 Here, $u_{\rm eff}$ is the disorder strength with $ u_{\rm eff}\;
\xi^2  \gg 1$ and $\zeta_i \in [0, 1]$ is a
random phase.  In the effective Hamiltonian Eq.~(\ref{eq.hamiltonian}),
the disorder term dominates the kinetic term and the classical ground
state of the system can be explicitly constructed \cite{Glatz01}. One
minimizes the cosine potential for each lattice site by letting
$p \phi_i = 2 \pi (\zeta_i +  n^0_i)$ with integer $n^0_i$.
The set of integers  $\{n^0_i\}$ is chosen in such a way
that the elastic term in Eq.~(\ref{eq.hamiltonian}) is minimized,
%
\begin{equation}
n^0_i = m + \sum_{i < j} \left[\zeta_{i+1} - \zeta_i \right]_G
\ \ .
\label{integersolution.eq}
\end{equation}
%
Here, $[ \zeta ]_G$ denotes the closest integer to $\zeta$, and $m$
is an integer parameterizing the infinitely many equivalent ground
states. Excitations of the ground state  change $n_i^0 \rightarrow
n_i^0 \pm 1$ for sites with $i_0 < i < i_0 + L_x /\xi $, they
bifurcate from one ground state characterized by $m = m_0$ to
another ground state with $m = m_0 \pm 1$. The potential energy
necessary for a bifurcation  at   position $i$ is according to
Eq.~(\ref{eq.hamiltonian})
%
\begin{eqnarray}
\Delta H(i)\! & = &\! {\Delta_0 \over 2 \pi^2} (2\pi /p)^2 \left\{ 1 \pm 2 \left( (\zeta_i -
\zeta_{i-1}) - \left[ \zeta_i - \zeta_{i-1} \right]_G \right) \right\}
\nonumber \\
& \equiv & \hbar v_{\rm eff} s_y g_{i}
\label{exitationenergy.eq}
\end{eqnarray}
%
with 
a random
$g_i \in [0,2]$. Defining the localization length $\xi_{\rm loc} = {p^2 K_{\rm eff}
\xi \over 2 \pi}$, one has $s_y = {1 \over \xi_{\rm loc}}$.

Quantum effects are due to the time derivative in the action
Eq.~(\ref{action.eq}) and give rise to tunneling between the ground
state and excited states in the presence of an external electric
field. Similar to the case of periodic pinning \cite{Maki77}, these
tunneling events are described by instantons.  In the following, we
describe how the action of a quadratic instanton can be calculated.

The action of a bifurcation with extension $L_y$ is just $\Delta H
(i_0) L_y/ v_{\rm eff}$. The action of a wall with constant $y$ and
length $L_x$ can be calculated by an analogous consideration if one
introduces a lattice of grid size $\xi$ in $y$--direction. As the
disorder is correlated in time direction, one needs not consider
random phases $\zeta_i$ and finds an action $\hbar s_y L_x$. Adding up
the contributions from all four walls of an instanton, one finds the
action of a rectangular $L_x \times L_y$--instanton \cite{NaGiDo03}
%
\begin{equation}
{\Delta S \over \hbar} = \left(s_x(i_0) + s_x(i_0 + L_x)\right) L_y
+ 2 s_y L_x, \,\,\,\ s_x(i)=s_yg_i.
\end{equation}
%
In a noninteracting electron system, a pair of sites with $g_i \ll 1$
corresponds to an electron with energy $-\epsilon_1 = - g_1 \Delta_0$
just below the Fermi level at position $i_0$, which can hop to an
unoccupied level with energy $\epsilon_2 = g_2 \Delta_0$ just above the 
Fermi level at position
$i_0 + L_x$. The translation between the language of noninteracting
elctrons and the bosonic language used in this calculation is
summarized in Table \ref{translation.tab}.

Typically, the two lowest $g_i$ in an interval of length $L_x$ are of the
order $1/L_x$, and the boundaries of a typical instantons will be
at positions with a small surface tension $s_x \approx s_y \xi/L_x$.
Taking into account the contribution of a dc external electric
field, the total action of a typical instanton is
%
\begin{equation}
{S(L_x,L_y) \over \hbar} = 2 s_y {\xi \over L_x} L_y  + 2 s_y
L_x - {2 e_0 E_0 \over p \pi v_{\rm eff} \hbar}  L_x L_y \ \ .
\label{quadratication_eq}
\end{equation}
%
Extremizing the action with respect to $L_x$, $L_y$ one finds \cite{NaGiDo03}
%
\begin{equation}
L_x(E) = \sqrt{2  \pi \over p \kappa e_0 E_0}\ , \ \ \ L_y = {  \pi \over p  \kappa
\xi e_0 E_0} \ \ .
\label{lengths.eq}
\end{equation}
%
The creation rate of these instantons is
%
\begin{equation}
P_{\rm random } \sim e^{- 2 L_x(E_0) / \xi_{\rm loc}}\ \ ,
\end{equation}
%
leading to the result Eq.~(\ref{quantum_creep.eq}) for the conductivity.

%
\begin{table}[htb]
\caption{\label{translation.tab} Translation between boson and electron language.}
\begin{ruledtabular}
\begin{tabular}{ll}
electron language & boson language \\
\hline
particle & kink \\
hole  & antikink \\
 $\rho(x)$ & $ {1 \over \pi} \partial_x \varphi$ \\
surface tension  $s_x(i_0)$  & energy $-\epsilon_1(i_0)$  \\
surface tension  $s_x(i_0+ L_x)$  & energy $\epsilon_2(i_0+ L_x)$  
\end{tabular}
\end{ruledtabular}
\end{table}
%

\medskip
Next we  consider an \emph{ac} field $E(t)$, which upon analytical
continuation $it \to \tau$ turns into a field $E(\tau)$. In
imaginary time, the electric field  has to obey the same
 periodic boundary
condition $E(\tau + \beta) = E(\tau)$ as other bosonic fields, e.g.
the displacement field $\varphi(\tau)$. This boundary condition is
respected by a discrete Fourier representation \cite{NeOr88}
%
\begin{equation}
E(\tau) = T \sum_{\omega_n} E(\omega_n) e^{- i \omega_n \tau}, \ \ \
\omega_n = {n 2 \pi k_B T \over \hbar} \label{matsubara.eq}
\end{equation}
with Matsubara frequencies $\omega_n$. A monochromatic external
field is hence described by $E(y) = E_0 \cos \tilde{\omega}_n y$,
where time is rescaled as $y = v_{\rm eff} \tau$ and frequency as
$\tilde{\omega}_n = \omega_n/ v_{\rm eff}$. In the end of our
calculation, we analytically continue Matsubara frequencies to
retarded real frequencies $i \tilde{\omega}_n v_{\rm eff} \to \omega
+ i \eta$.

%
%
\vspace*{1cm}
\begin{figure}[t]
\centerline{ \epsfig{file=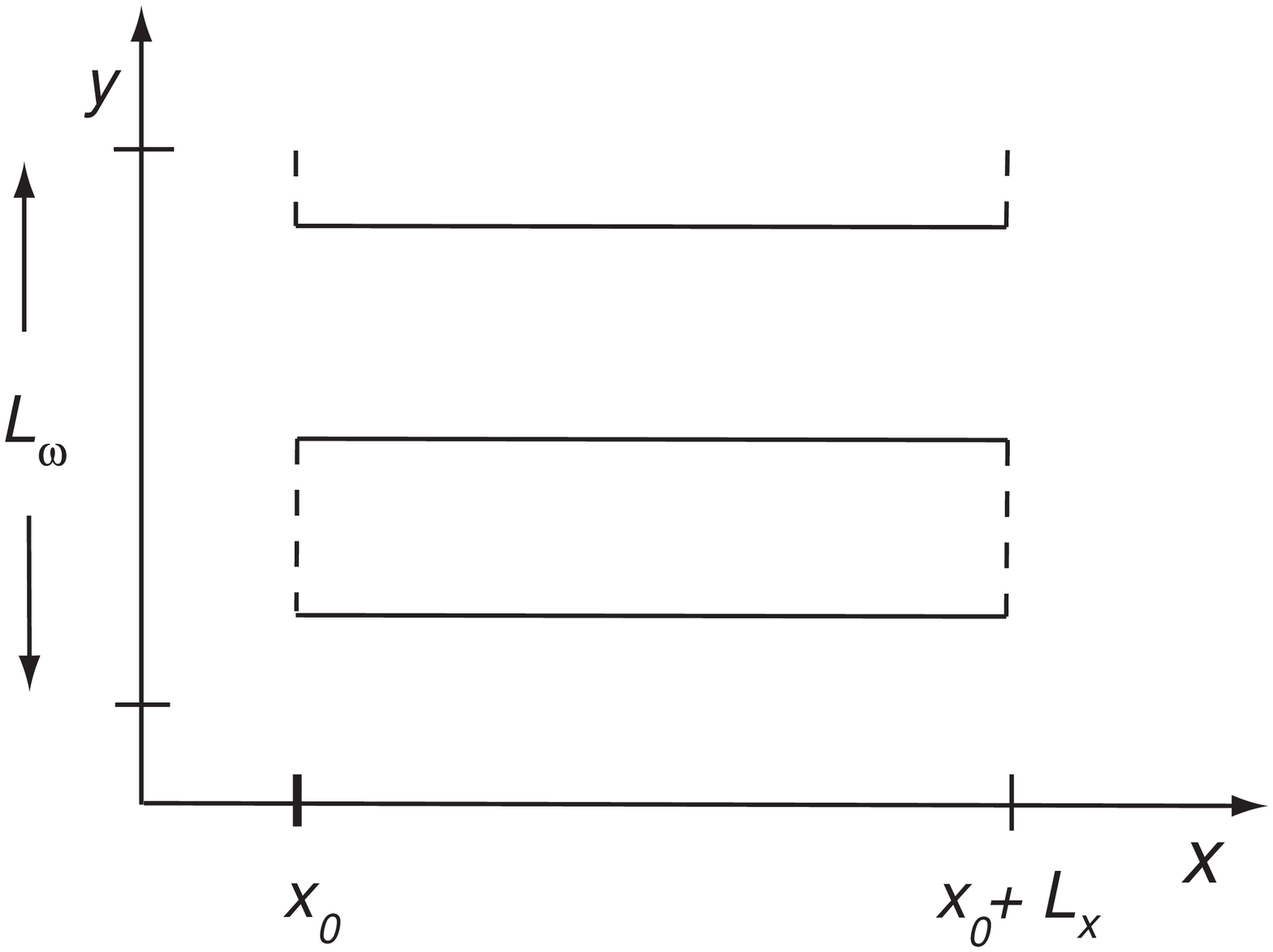,width=7cm}}
\caption{Hopping of kinks from position $x_0$ to $x_0 + L_x$ or
back (full lines) in a time interval of length $L_y$. The lines parallel
to the $y$--axis do not contribute to the action if the line tensions
$s_x$ are neglected.
}
\label{kinks.fig}
\end{figure}
%

Which type of  instantons determines the current in the presence of
an ac field with period $ L_\omega = 2 \pi / \tilde{\omega}_n $? In
the limit of a very small external electric field,
 the only external length
scale in the problem is the ac period $L_\omega$,
and the contribution of typical instantons to the linear response can be estimated by assuming
 that $L_y$ has to be of order $L_\omega$. A typical
instanton obtained from minimizing Eq.~(\ref{quadratication_eq}) with
respect to $L_x$ for a fixed $L_y = L_\omega$  and for vanishing $E_0$ has
$L_x = \sqrt{\xi L_\omega}$. For this solution, the current would be
proportional to $\exp[- {8 \pi \over p^2} (1+i)
\sqrt{\Delta_0/\hbar \omega }]$ and vanish nonanalytically for small
frequencies. However, according to the Mott-Halperin law, the true
frequency dependence should be proportional to \cite{MoHa68} $\omega^2
\ln^2(1/\omega)$ . We conclude that typical instantons
do not yield  the leading contribution to the current and that a discussion
of rare instantons is needed.

Indeed, besides typical instantons with $s_x \approx s_y \xi/L_x$,
there are rare  instantons with an exceptionally low $s_x(i) +
s_x(i+ L_x)$. Such a pair of sites $i$ and $i + L_x$ allows for the
hopping of a kink without changing the kink's potential energy much.
The potential energy difference $\epsilon_1 + \epsilon_2$ between two sites can become
arbitrarily small in sufficiently large samples. For the following
considerations, we will set it to zero in the sense that it is much
smaller than any other energy scale in the system. In the discussion
of the dc electric field, quantum fluctuations, i.e.~spontaneous
creation of typical instantons in the absence of an external field,
were unimportant.  For pairs of sites with exceptionally low surface
tensions, quantum fluctuations are important and have to be taken
into account. Here, the spontaneous formation of instantons
describes the physics of level repulsion \cite{NeOr88}. In our
approximation of vanishing $s_x$, the instanton action does not
depend on the extension $L_y$ in time direction any more, hence the
occurrence of single domain walls of length $L_x$ with constant $y$ is
possible. Such a domain wall describes the hopping of a kink across
the distance $L_x$, and its action is
%
\begin{equation}
S_{single}/\hbar = s_y L_x \ .
\end{equation}
%

To obtain the partition function for this tunneling degree of freedom, we
must sum over all possible domain wall  configurations in the
interval $L_\omega$. A configuration with three hopping events is displayed in
Fig.~\ref{kinks.fig}. Summation over all possible configurations yields
%
\begin{eqnarray}
Z(L_\omega) & = & \sum_{n=0}^\infty {1 \over n!} \prod_{i=1}^n
\left( \int_0^{L_\omega} {d y_i \over \xi K_{\rm eff} } \right)
e^{- n  s_y L_x}
\label{eq.partition}\\
& = & \sum_{n = 0}^\infty {1 \over n!} \left( {L_\omega \over \xi K_{\rm eff}}
e^{- s_y L_x} \right)^n \nonumber\\
& = & \exp\left({L_\omega \over \xi K_{\rm eff}} \ e^{- s_y L_x}
\right) \ . \nonumber
\end{eqnarray}
%
The integration measures in the $y_i$--integrals are normalized by $\xi K_{\rm eff}$ rather than by $\xi $ because the
short time cutoff  $\hbar / \Delta_0$ is determined by the high energy 
cutoff $\Delta_0$  and not by the short distance cutoff $\xi$. 
We note that the exponent of the outer exponential function is
positive, indicating a lowering of the ground state energy due to
frequent tunneling between the degenerate states.
The summation over all possible instanton configurations
describes the quantum mechanical effect of level repulsion.
Coupled energy levels repel each other and
are separated at least by
%
\begin{equation}
t_0 (L_x)=-\frac{\hbar v_{eff}}{L_{\omega}}\ln Z(L_{\omega})=
\Delta_0 \exp(- L_x/\xi_{\rm loc}) \ . \label{hoppingintegral.eq}
\end{equation}
%

 The
probability to have exactly one instanton (hopping of a  kink forth
and back) within the time interval $L_\omega$ is given by
%
\begin{eqnarray}
p_1 (L_x)  & = & e^{- 2 s_y L_x} / Z(L_\omega)
\label{eq.probability} \\
    & = & \exp\left(- {L_\omega \over \xi K_{\rm eff}} e^{- s_y L_x}
- 2 s_y L_x \right)\ . \nonumber
\end{eqnarray}
%
The optimal length $L_x$ for such an instanton is found by minimizing
the exponent in Eq.~(\ref{eq.probability}) with respect to the tunneling
length. We find
%
\begin{equation}
L_x = { 1 \over  s_y} \ln {L_\omega \over 2 \xi K_{\rm eff}} \ .
\end{equation}
%
Using this expression for $L_x$, we find the probability for having
exactly one instanton of length $L_\omega$
%
\begin{equation}
p_1  = e^{- 2} \; \left( {2 \xi K_{\rm eff}\over L_\omega} \right)^2 =  \left(
{ \tilde{\omega}_n \xi K_{\rm eff} \over e\; \pi}\right)^2  \ \ .
\label{instantonprobability.eq}
\end{equation}
%
This proportionality of the tunneling probability to frequency
squared is the essence of the Mott--Halperin conductivity
Eq.~(\ref{motthalperin.eq}).

With the knowledge of the probability
Eq.~(\ref{instantonprobability.eq}) for an instanton in resonance
with the external field, we can now set up a calculation of the ac
current. It  is calculated as a derivative $I(x,\omega_n) =-\hbar
{\delta \over  \delta   a(x,-\omega_n)} \ln Z$
of the partition function with respect to the vector potential
$a(x,\omega_n) = E(x,\omega_n) / \omega_n$. The field $\phi$ couples to
the vector potential via
%
\begin{equation}
S_E/\hbar = {e_0  \over \pi \hbar  } \int dx \; { 1 \over \beta v_{\rm eff}}\;
\sum_{\tilde{\omega}_n} a(x, -\tilde{\omega}_n)(- \tilde{\omega}_n)
\phi(x,\tilde{\omega}_n) \ ,
\end{equation}
%
where  $\beta$ denotes the inverse
temperature. Hence, the current is given by
%
\begin{equation}
I(x,\tilde{\omega}_n) = - e_0 \;   \tilde{\omega}_n   v_{\rm eff} {
\int D[\phi]  \phi(\tilde{\omega}_n) e^{- S[\phi] / \hbar} \over
\int D[\phi] e^{- S[\phi]}} \ .
\end{equation}
%

In the low energy regime, we do not perform the full functional
integral over $\phi$ in order to evaluate the partition function.
Instead, we sum over the relevant tunneling degrees of freedom. We
label such a degree of freedom by the position $i_0$ of the first
weak link and by the distance $L_x$ between the first and the second
weak link. The field $\phi(i_0)\; = \; \phi(i_0 +1) \; = ...=
\phi(i_0 + L_x)$ takes the values $\phi_0$ and  $\phi_0 + 2 \pi /p$,
respectively. Furthermore, we make use of the self averaging
properties of the current. Instead of averaging it over the position
$x$ in the system and let all different types of TLSs contribute to
it, we calculate the contribution of one TLS and average over the
parameters $L_x$, $s_x(i_0)$, and $s_x(i_0 + L_x)$. So far we
considered instantons with vanishing surface tension $s_x=0$. We now extend these considerations to instantons  with surface energies
smaller than $\hbar \omega$, i.e.  we are concentrating on sites
with $g_i < K_{\rm eff} \xi \tilde{\omega}_n$, where
$s_x(i_0)=s_yg_1,\, s_x(i_0 + L_x) = s_yg_2$.
The probability that the position $x$, for which
we want to evaluate the current, is inside an active instanton is
$L_x / \xi$ times the probability for finding a weak tunneling link
at a given site.
 In this way, we obtain for the average current
%
\begin{eqnarray}
\hspace*{-.8cm} \langle I \rangle (\tilde{\omega}_n) & = &- e_0
\tilde{\omega_n} v_{\rm eff} \int {dg_1} {dg_2} { d L_x \over \xi} \
{ L_x \over \xi }\  p_1(L_x)  \cdot  \nonumber \\[.3cm]
& &  \cdot  \int {d y_1\over \xi K_{\rm eff}}{ d y_2
\over  \xi K_{\rm eff}}\ \phi_0(\tilde{\omega}_n; y_1, y_2) \
 e^{- S_E[\phi_0]}
\end{eqnarray}
%
Here, we evaluate the current under the approximation that there is
exactly one instanton (two tunneling events at times $y_1$ and
$y_2$) in the interval of length $L_\omega$.
The two $g_i$
integrals contribute a factor $(K_{\rm eff} \xi \tilde{\omega}_n
)^2$ as we neglect the dependence of $S_E$ on the $g_i$.  The integral over $L_x$ is evaluated by the saddle point
method just taking into account instantons of the optimal length
$L_x = {1 \over
  s_y} \ln{L_\omega \over 2 \xi K_{\rm eff}}$. As $p_1$ is a function of $L_x / \xi_{\rm loc}$,
  we have to use the same variable in the integration measure to perform a 
  saddle point approximation and obtain
  a factor $\xi_{\rm loc} / \xi $ from this transformation.
If $y_1$ is the time of
the first tunneling event and $y_2$ the time of the second tunneling
event, we define the new variables $\tilde{y} = (y_1 + y_2)/2$ and
$L_y = y_2 - y_1$. The Fourier transform of the displacement field
$\phi$ for such an instanton is given by
%
\begin{equation}
\phi(\tilde{\omega}_n;L_y,\tilde{y}) = {1 \over \pi} \ e^{i
  \tilde{\omega_n} \tilde{y}}\ \sin{\tilde{\omega}_n L_y \over 2} \ .
\end{equation}
%
For this field configuration, the coupling to the external electric field
contributes the action
%
\begin{eqnarray}
S_E(L_y,\tilde{y})/ \hbar
& = &  { e_0 E_0 L_x \over \pi \hbar v_{\rm eff}} {2 \over\tilde{ \omega}_n}
\sin (\tilde{\omega}_n \tilde{y}) \ \sin (\tilde{\omega}_n L_y /2) \nonumber \  .\\
\end{eqnarray}
%
The barrier size  is not fixed by the electric field as in the dc limit, and
one has to consider both  forward and backward jumps as for
the standard thermally assisted flux flow argument \cite{AnKi64}.
Hence, the probability for the instanton to be in phase with the external
field is  given by $2 \sinh [
  S_E(L_\omega,L_x)/ \hbar ]$.  Expanding to linear order in $S_E$ and
integrating over $L_y$ and $\tilde{y}$, we find for the current
%
\begin{eqnarray}
\langle I \rangle (\tilde{\omega}_n) & = &- e_0 v_{\rm eff} \tilde{\omega}_n
\ \ ({\tilde{\omega}_n \xi K_{\rm eff} \over e \; \pi})^2 \ \ (\tilde{\omega}_n
\xi K_{\rm eff})^2 \ \ {L_x \over \xi} {\xi_{\rm loc} \over \xi} \nonumber \\
& &  \hspace*{-0.2cm}
\int_0^{L_\omega /2} \! \! \! \! { d \tilde{y} \over \xi K_{\rm eff}} \int_0^{L_\omega } \! \! \! \! {d
  L_y \over \xi K_{\rm eff}} \ \phi(\tilde{\omega}_n;L_y,\tilde{y}) \ {2
  S_E(L_y,\tilde{y})\over \hbar} \nonumber \\
& = &-  {8 \over
  e^2} \; {e_0^2 \over \hbar} \; \xi_{\rm loc} \;L_x^2 \;  \; \tilde{\omega}_n^2 \; K_{\rm
    eff}^2\ E_0
 \label{finitefrequency.eq}
\ \ .
\end{eqnarray}
%
After the analytical continuation $\tilde{\omega}_n v_{\rm eff} \to
-i \omega$ the real part of the conductivity agrees with formula
Eq.~(\ref{motthalperin.eq}) up to a numerical factor. As both
$\xi_{\rm loc}$ and $L_x$ are proportional to $K_{\rm eff}$, the
conductivity contains a factor of $K_{\rm eff}^5$ in agreement with
the result \cite{FeVi81,Fogler02}.
\\

\setcounter{equation}{0}

\section{Description by Bloch equation}

The instanton calculation presented in the last section needs to
be improved upon in two respects: first, the classical level separation
parameterized  by the $g_i$ was not fully taken into account, and  second,
nonlinear correction in the strength of the external field
are not considered yet. To achieve these goals, we use the concepts
developed in the last section in a real time quantum mechanical calculation.
Instantons in the
imaginary time formalism correspond to the hopping of kinks from one
level just below the chemical potential to another level just above
the chemical potential. The localized states of a 1d disordered system can be modeled
by an ensemble of these TLSs, and  the average properties of
the system can be calculated by averaging over the parameters of the
TLSs.

We consider a  TLS with spatial extension $L_x$ and on site
energies $-\epsilon_1$ and  $\epsilon_2$ with $\epsilon_i = g_i \Delta_0$. 
In a system of noninteracting electrons, the negative energy $-\epsilon_1$ corresponds to a particle below the Fermi level, 
and the positive energy  $\epsilon_2$ to an unoccupied site above the Fermi level.
The two sites are coupled by
a distance dependent hopping integral $t_0(L_x)$ according to
Eq.~(\ref{hoppingintegral.eq}).
Such a TLS is described by the
Hamiltonian $\hat{H}=\hat{H}_0 + \hat{H}_E$ with
%
\begin{equation}
\hat{H}_0 =
{1 \over 2} (\epsilon_1 + \epsilon_2) \sigma_z -  t_0(L_x)\; \sigma_x  ,
\hat{H}_E =
 { 1 \over 2} \ e_0 L_x E_0 \cos \omega t \ \sigma_z \ .
 \end{equation}
%
The position of the tunneling kink is
measured by the operator $\hat{x} = {1 \over 2} L_x \sigma_z$, and the
current operator is given by $\hat{I} =
\rho(\epsilon_1,\epsilon_2,L_x) e_0 \dot{\hat{x}} $.  Here,
%
\begin{equation}
\rho(\epsilon_1,\epsilon_2,L_x)={1 \over \xi} {d \epsilon_1 \over
  \Delta_{0}} {d \epsilon_2 \over \Delta_{0 }} {d L_x \over \xi}
\label{density.eq}
\end{equation}
%
denotes the spatial density of TLSs with given parameter values.
 $\hat{H}_0$ is
diagonalized by the unitary transformation
%
\begin{equation}
 \hat{H} \to    \exp(i \varphi \sigma_y
/2) \hat{H} \exp(- i \varphi \sigma_y /2)
\end{equation}
%
with $\varphi =
\arctan { 2 t(L_x) \over \epsilon_1 + \epsilon_2}$.  In the new basis,
$\hat{H}_0$ corresponds to a static field in $z$--direction, and
$\hat{H}_E$ to an oscillating field with $x$--component proportional
to $\sin \varphi$ and $z$--component proportional to $\cos \varphi$.

In principle, the transformed  Hamiltonian should now be solved in a
nonequilibrium setup in a dissipative environment. In general,
this type of problem is difficult to deal with in full
generality \cite{Weiss99}. However, the problem  simplifies if
one does not treat an individual quantum system but averages over
a whole ensemble instead. Such an ensemble of TLSs or spins
interacting with an oscillatory electric field and subject to relaxation
processes can be described by  Bloch equations \cite{CoDiLa77}. We denote the
ensemble polarization of the TLSs in the transformed basis by the
pseudospin vector
$ \underline{p}$. The current $I$ is then proportional to the
the pseudospin component in
$y$--direction,
%
\begin{equation}
I = -{1 \over 2} p_y
\rho(\epsilon_1,\epsilon_2,L_x) e_0 L_x {t_0(L_x) \over \hbar} \ .
\end{equation}
%
The pseudospin polarization $\underline{p}$ of the TLSs
follows the Bloch equation
%
\begin{equation}
{d \over d t}\; {\underline{ p}} = - {1 \over \tau_0}\;
(\underline{p} -  \underline{ p}_0)
 + \alpha\; {\underline{ p}} \times \underline{E} \ \ .
\label{bloch.eq}
\end{equation}
%
Here, $\underline{E} = \left[ E_0 \sin (\varphi) \cos( \omega t),0, {2
\Delta \over e L_x} + E_0 \cos(\varphi) \cos(\omega t)\right]$,
$\underline{p}_0 = [0,0,-1]$, $\Delta = \sqrt{(\epsilon_1^2+
\epsilon_2^2)/4 + t_0^2(L_x)}$, and $\alpha = e_0 L_x / \hbar$.
Inelastic processes are described by the phenomenological damping
constant $\tau_0$. In the absence of an external electric field, the
pseudospin relaxes due to this damping to its equilibrium value $\underline{ p}_0$,
i.e. the particle is in a superposition of states localized at $i_0$
and $i_0 + L_x$. At temperatures much lower than the hopping integral
$t_0(L_x)$, the dissipative bath cannot destroy this coherent
superposition and localize the particle, as the localized states have a
higher energy than the symmetric linear combination.

\begin{figure}[t]
\centerline{ \epsfig{file=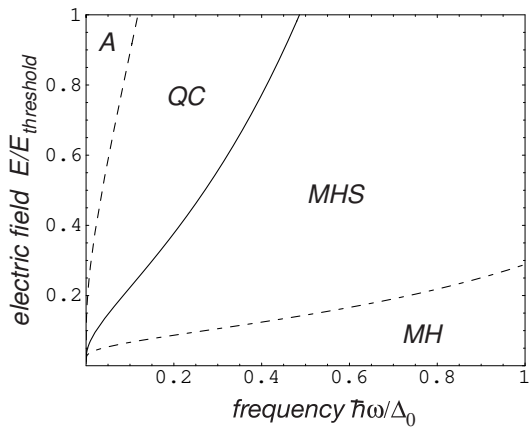,width=7cm}}
\caption{Different transport regimes in a disordered CDW or LL with
  dissipation as a function of field strength in units of $E_{\rm
    threshold} = \Delta_0 /p e_0 K_{\rm eff} \xi_{\rm loc}$ and
  frequency: adiabatic quantum creep in A, quantum creep with
  increased current noise in QC, ac conductivity following a
  Mott-Halperin law in MH, Mott-Halperin law with saturation for large field strengths in MHS.}
\label{crossover.fig}
\end{figure}

The solution of Eq.~(\ref{bloch.eq}) is described in detail in reference
\cite{CoDiLa77} and we do not reproduce it here. We find that
to order $O(E_0^3)$, one type of TLS contributes to the conductivity
%
\begin{eqnarray}
\sigma_{\rm TLS}(\omega;\! \epsilon_1,\! \epsilon_2,\! L_x\!) &=&
\rho(\! \epsilon_1,\! \epsilon_2, \! L_x\!)  \; {e_0^2 L_x^2
  t^2(L_x) \over 2 \hbar^2 \Delta}  \cdot
\nonumber \\
& &
\hspace*{-3cm}
 \cdot {- i \over \omega - {2 \Delta
    \over \hbar} - i {\hbar \over \tau_0}}\left[ 1 - {
   1 \over \hbar^2  \Delta^2} \; { (e_0 L_x E_0 t_0(L_x))^2
        \over (\omega - {2 \Delta \over \hbar})^2
    + {1 \over \tau_0^2}}\right] \! .
\label{TLSconductivity.eq}
\end{eqnarray}
%
The reduction of the linear conductivity becomes effective for strong
ac fields, when both states of the TLS are occupied with comparable
probability.  In order to calculate the conductivity of the disordered
sample, we integrate over all possible parameter values $\epsilon_1$,
$\epsilon_2$, and $L_x$ and obtain the final result
\begin{equation}
{\rm Re} \sigma_{\rm ac} (\omega)  \!= \! \sigma_0
{\pi \over 4}  L^2_x(\omega)  (\hbar
\omega \kappa)^2  ( 1\! -\! 2\! {e_0^2 L_x(\omega) ^2 E_0^2 \over
  \hbar^2/\tau_0^2})
\label{acconductivity.eq}
\end{equation}
%
with the optimal tunneling length given by
%
\begin{equation}
L_x(\omega) = \xi_{\rm loc} \ln {2 \Delta_0 \over \hbar \omega} \ \  .
  \label{tunnellength.eq}
\end{equation}
%
The linear part of Eq.~(\ref{acconductivity.eq}) is proportional to
$K_{\rm eff}^5$ and agrees with the result of Fogler \cite{Fogler02} .
This linear conductivity describes the response of a disordered 1d system in 
region MH of Fig.~\ref{crossover.fig}.
For  an unscreened Coulomb interaction,  in Eq.~(\ref{acconductivity.eq}) one factor of $\hbar \omega$ has
to be replaced by  \cite{S81} $e_0^2/\epsilon L_x(\omega)$, where $\epsilon$ is
the dielectric constant of the system.

When $e_0 E_0 L_x(\omega) \approx \hbar / \tau_0$, higher order terms
become important and the ac current will saturate as a function of
$E_0$.  The value $E_s$ of the electric field where the current saturates can 
be estimated from Eq.~(\ref{acconductivity.eq}) as
%
\begin{equation}
E_s =  {\hbar \over e_0 \tau_0 L_x(\omega)}  \ \ .
\end{equation}
%
As the nonlinear conductivity is defined as the ratio of current and electric field,
in the saturation regime one obtains 
%
\begin{equation}
\sigma_{\rm ac} (\omega, E > E_s) = \sigma_0 {p \over 8} L_x(E)^2 {L_x(\omega) \over \xi} {\hbar \over \tau_0 \Delta_0} (\hbar \omega \kappa)^2 \ \ .
\label{satcond.eq}
\end{equation}
%
The region in $\omega$-$E$ space where the  
nonlinear conductivity Eq.~(\ref{satcond.eq}) can be observed is labeled MHS in Fig.~\ref{crossover.fig}.

%
%
\vspace*{1cm}
%

\setcounter{equation}{0}

\section{Discussion}

How does the linear conductivity Eq.~(\ref{acconductivity.eq})
connect to the creep current in strong fields?
The calculation of the nonlinear dc conductivity
involves the optimal length scale $L_x(E)$ in
Eq.~(\ref{lengths.eq})  for tunneling processes.  The crossover from ac to dc
conductivity takes place when the two length scales $L_x(\omega)$ and
$L_x(E)$ match, i.e. for a crossover frequency
%
\begin{equation}
\omega_{\times} = {\Delta_0 \over \hbar}  e^{- L_x(E_0) / \xi_{\rm loc}} \ .
\end{equation}
%
For $\omega \approx \omega_\times$, the magnitude of the creep current
$I_{\rm creep} \sim \exp(-2L_x (E)/ \xi_{\rm loc})$ agrees with the
magnitude of the ac current described by the conductivity  Eq.~(\ref{acconductivity.eq}), as the
$\omega^2$ term in Eq.~(\ref{acconductivity.eq}) matches the exponential
dependence on field strength of $I_{\rm creep}$. Then, the expression
Eq.~(\ref{satcond.eq}) turns into
%
\begin{equation}
\sigma(E) = \sigma_0 {p \over 8} {\hbar \over \tau_0 \Delta_0} {L_x(E)^3 \over \xi^3}  e^{- 2 L_x(E)/ \xi_{\rm loc} }\ \ ,
\label{weiss.eq}
\end{equation}
%
providing us with an estimate of the prefactor of the exponential factor describing dc creep.
Identifying $\hbar/\tau_0 \Delta_0$ as a dimensionless measure for the 
dissipation strength, this estimate agrees with a more sophisticated 
calculation, in which a TLS is coupled to phonons with an Ohmic spectral 
function \cite{Rosenow04}.

 In the crossover
region $\omega \approx \omega_\times$
between regimes QC and MHS in Fig.~\ref{crossover.fig}, 
there are two different types
of TLS contributing to the current.  In the ground state of a typical
TLS with bare energy difference $\epsilon_1 + \epsilon_2 \approx
\Delta_0 \xi/L_x(E) \gg t(L_x)$, most of the charge is localized at
one of the levels. Under the influence of an external field, the
charge hops {\em irreversibly} from one level to the other , as the
hop is generally accompanied by an inelastic process. On the other
hand, the ground state of a TLS with exceptionally low bare energy
separation $\epsilon_1 + \epsilon_2 \leq t(L_x)$ is the even parity
combination of wave functions centered around the individual levels,
and absorption of a photon excites the TLS to the odd parity state.

In the quantum creep regime, the time dependence of the current is calculated
using the time dependent field in the formula for the dc current $I_{\rm creep}$.
The adiabatic regime (region A in Fig.~\ref{crossover.fig}) is reached for frequencies smaller than the dc
hopping rate $I_{\rm creep}/e_0$.  While the average current is
stronger than the current noise in the adiabatic regime, the current
noise is stronger than the current for $I_{\rm creep} /e_0 < \omega <
\omega_\times$.

In summary, we have discussed the crossover from a nonlinear creep
current in a static electric  field to the linear ac response in 1d disordered
interacting electron systems.  While  the
 linear ac conductivity is described by a
generalized Mott-Halperin law, for stronger fields one finds
a reduction of this linear conductivity 
as both states of a TLS are occupied with comparable probability.
 The crossover between nonlinear ac  conductivity and
dc creep current occurs when the spatial extension of TLSs matches
the length scale for tunneling of 
kinks.

\acknowledgments We thank T.~Giamarchi, S.~Malinin,
D.~Polyakov, and B.I.~Shklovskii for useful discussions
and the SFB 608 for support.

\end{document}